# Influence of hydrogen vacancy interactions on natural and artificial ageing of an AlMgSi alloy

Guillaume Hachet[(*)] and Xavier Sauvage[(*)]

Groupe de Physique des Matériaux, Normandie Université, UNIROUEN, INSA Rouen, CNRS, 76000 Rouen, France.

[(*)] Corresponding authors: guillaume.hachet@univ-rouen.fr, xavier.sauvage@univ-rouen.fr

**Abstract**: The influence of hydrogen on the structural evolutions of an Al-Mg-Si alloy during natural and artificial ageing was investigated experimentally. The aim of the study was a better understanding of interactions between hydrogen and crystalline defects and especially vacancies. Experimental data demonstrate that during natural ageing in hydrogen environment, the hardening response is delayed. This is attributed to a slower recovery of excess vacancies linked to a lower mobility. To confirm and quantify the influence of hydrogen on the vacancy migration energy, artificial ageing was carried out in conditions where the vacancy concentration is constant. Hence, long-time annealing treatments were carried out to investigate the influence of hydrogen on the coarsening of rod-shaped precipitates. Using transmission electron microscopy and atom probe tomography, it was demonstrated that the precipitate volume fraction and composition are unchanged under $H_2$ atmosphere but the coarsening kinetic is significantly reduced. This leads to a delayed softening, in good agreement with theoretical estimates. Thus, even a low concentration of hydrogen in solid solution significantly affects the mobility of alloying elements in the aluminium matrix. This is the result of hydrogen-vacancy interactions that lead to an increase of the vacancy migration energy. Based on classical coarsening theories, it was possible to demonstrate that this increase is of about 5% for a concentration of hydrogen close to the vacancy concentration.

## 1. Introduction

Hydrogen became a popular candidate as an alternative green energy carrier. However, with a small size, a high mobility and strong interactions with crystalline defects, this element is also well known for premature failures attributed to the so-called hydrogen embrittlement of metallic alloys [1,2]. Several models have been proposed in the literature to describe the underlying physical mechanisms (see for details [3–6]). The most popular ones are the Hydrogen Enhanced Decohesion (HEDE) [7–9] based on the hydrogen induced reduction of cohesive interfaces; the Hydrogen Enhanced Local Plasticity (HELP) [10–13] based on hydrogen enhanced plasticity localisation and dislocation mobility; the Super Abundant Vacancies (SAV) [14–17] based on hydrogen enhanced vacancy formation energy; and the embrittlement induced by the local formation of hydrides [18]. A common feature of these models is the reduction of the activation energy of one physical process (dislocation mobility, vacancy formation, etc.) due to hydrogen, and thus it has been formulated using the "defactant concept" [19]. In all cases, it involves the interaction of hydrogen with defects and even if the topic has been investigated since a long time, there is a lack of direct experimental quantification of such interactions. This is due to the intrinsic difficulty of hydrogen imaging and quantification by conventional techniques [20,21], therefore this issue has been mainly treated so far using atomic scale modelling approaches. Nevertheless, there is a domain in physical metallurgy where crystalline defects play a major role, namely precipitation [22–24]. At all stages (nucleation, growth and coarsening), the vacancy concentration and mobility tune the kinetic (in case of a classical vacancy-controlled diffusion mechanism) and interfaces the thermodynamics (nucleation barrier, precipitate morphology, driving force for coarsening).

The aim of the present work was to investigate experimentally the influence of hydrogen on the precipitation mechanisms in a model aluminium alloy for a better understanding of interactions between hydrogen atoms with these defects (vacancies and heterophase interfaces). A precipitate hardening Al-Mg-Si alloy was selected since the precipitation sequence is well known in such alloys [21-25, 34] and also because they are considered as possible candidates for high-pressure hydrogen storage cylinders [25]. After homogenisation and quenching, the Mg and Si super saturated solid solution quickly decomposes leading to the formation of clusters during natural ageing near room temperature [26]. During artificial ageing at higher temperature (typically in a range of 160 to 200°C), needle shaped metastable phases (called $\beta''$ and $\beta'$) nucleate and grow, leading to a significant hardness increase. Besides, it has been shown that in this system, precipitate/matrix interfaces are strong trapping site for tritium [27]. Thus, presumably hydrogen atoms may behave similarly and affect the precipitation kinetic. Moreover, it has been clearly demonstrated using *ab initio* calculations that hydrogen atoms also strongly interact with vacancies in aluminium [28–30]. Therefore, to experimentally confirm these features and quantify these interactions, an Al-Mg-Si alloy was aged in two different conditions: under air and under $H_2$. Transmission Electron Microscopy (TEM) observations and Atom Probe Tomography (APT) analyses were then carried out to compare the precipitate number density, morphology, composition and mean size. Finally, using classical theories for coarsening and precipitate hardening, a relationship between microstructural changes during ageing and the resulting micro-hardness was established to quantify the influence of solute hydrogen on interfacial energy and on atomic mobility.

**2. Experimental procedures**

The investigated material is an AA6201 with the following composition (wt.%): 0.81% Mg-0.79% Si, Al balance. Disc shaped samples (diameter 20mm, thickness 1mm) were solutionised at 540°C during 1h, water quenched and then naturally aged in air at 20°C. To evaluate the impact of hydrogen on the natural ageing response, some samples were put in contact with a 1M NaOH solution at 20°C during 5h after quenching and then further aged in air at 20°C. Before inserting the alloy in the NaOH solution, samples were quickly (few minutes) mechanically grinded using SiC foil paper with a particle size of 8 μm to remove the oxide layer grown during the solution heat treatment. Aqueous solution containing NaOH is aggressive towards aluminium and its oxide, it prevents the formation of a passive layer and leads to H incorporation in the alloy [4,31]. To evaluate the influence of hydrogen on the artificial ageing response, heat treatments were carried out at temperatures ranging from 140°C to 200°C in air or under 1 bar of $H_2$. Prior to these treatments, samples were solutionised, quenched, mechanically grinded and naturally aged during a week in air. They were introduced in a chamber that was evacuated, filed with 1bar $H_2$ and then heated up to the required temperature (heating time shorter than 15min). The diffusion coefficient of H ($D_H$) in FCC aluminium at such temperatures is 0.72 $10^{-11}$ and 5.4 $10^{-11}$ $m^2s^{-1}$ respectively [32]. Then, the time $t$ needed to achieve an effective diffusion length $L$ corresponding to half of the disc thickness is about 45 min at 160°C ($D_H$ =1.5 $10^{-11}$ $m^2s^{-1}$) and only 12 min at 200°C (with L = $(6 D_H \times t)^{1/2}$). Artificial ageing (AA) was carried out during 24h, so most of microstructural evolutions in samples aged under $H_2$ occurred in conditions where both the vacancy concentration and the amount of hydrogen in solid solution were relatively constant.

The mechanical behaviour after ageing was evaluated thanks to micro-hardness measurements performed with a Future tech FM7 device at room temperature. The micro-hardness values presented in this study are the average of at least 6 indents obtained with a micro Vickers diamond indenter using a load of 300g and a dwell time of 10 s.

Microstructure evolutions were observed by TEM with a JEOL-ARM200F microscope operated at 200kV. Both conventional TEM and Scanning TEM (STEM) were carried out. STEM High angle annular dark field (HAADF) images were recorded with collection angles ranging from 67 to 250 mrad. Electron transparent specimens were prepared with a twin-jet electro-polisher (TENUPOL 5 from Struers®) using a mixture of 30%$HNO_3$-70% $CH_3OH$ (%vol) at -30°C. Final thinning was carried out by low-energy ion milling conducted with a GATAN® Precision Ion Polishing System. Precipitates and matrix compositions were measured by APT with a CAMECA® LEAP 4000HR instrument. APT samples were prepared by classical electropolishing methods at ambient temperature [33,34]. They were then field-evaporated at 50K with electric pulses (pulse repetition rate 200 kHz, pulse fraction 17%). Data processing was done with the "IVAS 3.8.0" software from CAMECA®.

## 3. Effect of hydrogen on the natural ageing of AlMgSi

Few minutes after quenching, the microhardness is only 48±2 HV and it progressively increases up to 69±2 HV during ageing at room temperature in air (Fig. 1). This typical behaviour is in good agreement with previous studies on natural ageing mechanisms in AlMgSi alloys [35–37]. The recovery of excess vacancies leads to the quick formation of solute clusters and a fast hardening within few hours and then, a microhardness saturation is finally reached after few days. Remarkably, the alloy aged 5h in NaOH exhibits a microhardness significantly lower than that of the alloy aged in air during the same time (64±2 HV against 69±2 HV, Fig. 1). This indicates that the clustering kinetic was most probably slowed down by solute hydrogen. When the alloy is aged 5h naturally in air, the slow hardness increase corresponds to the diffusion of magnesium within the solid solution, which is controls the formation of co-clusters of both Si and Mg [35]. Thus, hydrogen may impact this diffusion. However, after several additional hours at room temperature in air, the micro-hardness further increases and catches up the hardness of the material naturally aged only in air.

This suggests that hydrogen atoms quickly desorb from the alloy and do not significantly affect clustering after 10 h.

The delay in clustering could either be due to a reduction of the vacancy concentration (by increasing the vacancy formation energy) or vacancy migration (by increasing of the vacancy migration energy) [38]. *Ab initio* calculations have shown that interactions between hydrogen atoms and vacancies are attractive in aluminium (interaction energy between -0.25 eV [29] and -0.33 eV [28]), indicating that hydrogen atoms most probably stand in the vicinity of vacancies. But, *ab initio* calculations have also shown that hydrogen can affect the vacancy mobility both ways (increases in Ni [39] and decreases in Cu or Pd [40]). Since H atoms may be located on octahedral or tetrahedral sites in FCC metals [29], their exact position in the lattice may influences diffusion mechanisms. In addition, solute atoms (Mg or Si) in the alloy of the present study might bring further complexity due to possible vacancy-solute interactions. The influence of H on the vacancy formation energy seems more systematic and it has been shown by X-ray diffraction experiments that it is usually decreased in FCC metals [17,41]. This is also one of the main hypotheses of the SAV embrittlement model [41]. Then, the delayed hardening during natural ageing in contact of NaOH is most probably the result of a decrease of vacancy migration leading to a lower mobility of Mg and Si solutes.

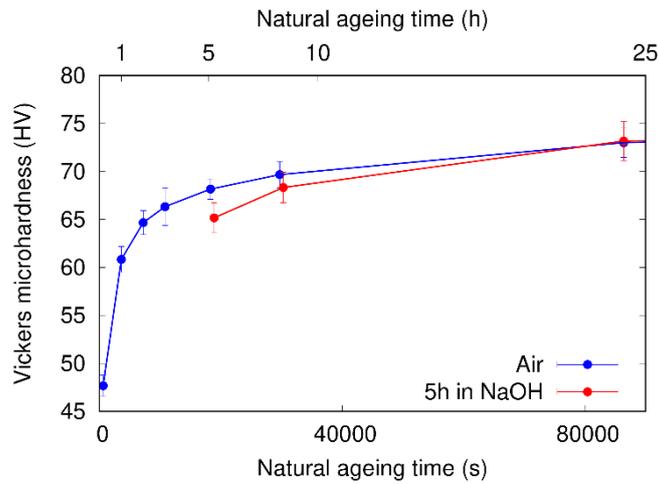

Figure 1: Vickers micro-hardness evolution as a function of the natural ageing time in air (blue) and after 5h in a NaOH solution (red).

In such a case, the shift between hardening curves in fig. 1 can be directly correlated to a change of the diffusion coefficient. After 5h in NaOH, the hardness is similar to that measured when the alloy is aged in air during 2.4h. The average diffusion distance $\overline{L^i}$ of solutes (Mg or Si) can be written as:

$$\overline{L^i} = \sqrt{6D_{eff,s}^i t}, \quad (1)$$

with $t$ the time, $D_{eff,S}^i$ the effective diffusion coefficient of solutes and $i = A\ or\ H$ for natural ageing conditions in air and in NaOH, respectively. After 5h in NaOH, the hardness is similar to the hardness measured when the alloy is aged in air during 2.4h, thus $\overline{L^H}(t=5h) = \overline{L^A}(t=2.4h)$ leading to $D_{eff,S}^H \approx 0.5 \times D_{eff,S}^A$. This simple estimate means that hydrogen induces a reduction of about 50% of the atomic mobility during natural ageing. Since it may take some time for hydrogen to diffuse into the core of the sample once in contact with NaOH, the influence of H on the mobility of vacancies might be even larger. Natural ageing occurs however in a situation where the vacancy concentration is not constant and not at the equilibrium (recovery of excess vacancies captured by quenching), thus to better quantify the interaction of H with vacancies, additional artificial ageing experiments were carried out.

**4. Impact of hydrogen on the artificially aged AlMgSi alloy**

*4.1 Influence of hydrogen on precipitate hardening*

The micro-hardness of the AlMgSi alloy artificially aged during 24h at temperatures between 140 and 200°C is plotted in fig.2. The micro-hardness lies between 120 and 75 HV with a maximum near 160°C and the lowest values was recorded at the highest ageing temperatures. The precipitate volume fraction depends on the temperature (due to a solubility limit change), but beyond 160°C as shown in previous studies [29] the material is overaged and precipitate coarsening is well advanced. The samples aged at 160°C under air and $H_2$ exhibit very similar hardness (118±3 HV and 117±3 HV, respectively). However, a significant difference is exhibited at 200°C (76±3 HV and 92±3 HV, respectively). Thus, hydrogen seems to affect the evolution of precipitates mostly in the coarsening regime, and since the micro-hardness is higher for the sample aged under $H_2$, it seems that the kinetic was slowed down by hydrogen in solution. To clarify this point, APT analyses and TEM observations were carried out.

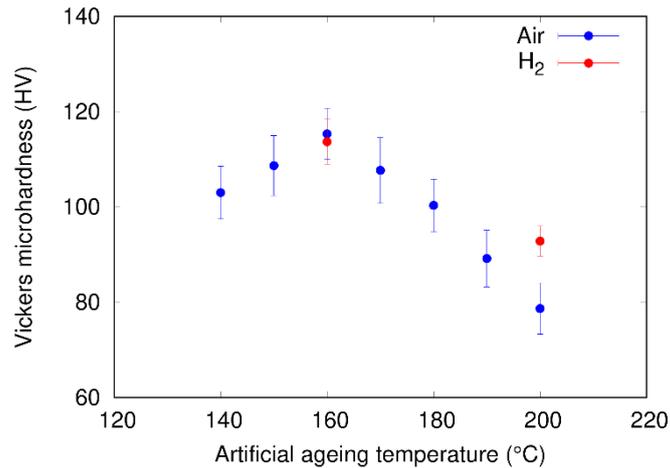

Figure 2: Vickers micro-hardness of the AlMgSi alloy as a function of the AA temperature (duration 24h) carried out in air and under $H_2$ environment.

After AA at 160°C in air or in H₂, a high density of nanoscaled precipitates is exhibited in both materials and no significant difference could be detected (fig.3).

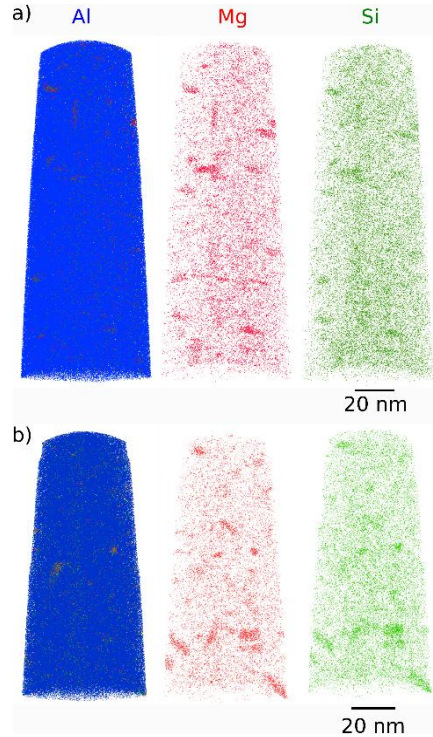

Figure 3: Three dimensional reconstructions of volumes analysed in the AlMgSi alloy aged 24h at 160°C in air (a) and under H₂ (b). For each state the same volume is represented three times: with all atoms (Al blue, Mg blue, Si green), only Mg and only Si to clearly exhibit the precipitates.

After 24h at 200°C (fig. 4), precipitates are significantly larger indicating that the coarsening regime is well advanced. Matrix and precipitate compositions extracted from these data are reported in table 1. A small amount of Al is detected in precipitates and they contain mainly Mg and Si as expected. The Mg/Si ratio estimated from APT data (in %at) is about 2.1 in both samples, corresponding most probably to the $\beta'$ phase [42] even if is slightly above the expected ratio [43]. Matrix compositions are also very similar with only about 0.02 at.% of Mg and 0.10 at.% of Si left in solid solution in both samples, indicating that hydrogen does not affect significantly the solubility limit. The volume fraction $f_v$ was estimated considering the mass balance [44]:

$$C_j^m(1-f_v) + C_j^p \alpha f_v = C_j^0\big(1 - f_v(1-\alpha)\big), \quad (2)$$

with $\alpha$ the ratio of atomic volume between the matrix and the precipitate ($\alpha = 0.865$ [44]), $C_i^0$ the initial atomic fraction of solute $j$, $C_j^m$ the concentration of solute $j$ in the matrix, and $C_j^P$ the mean concentration of solute *j* in the precipitates, with *j = Mg* or *Si*. Due to the possible overlap between $Si_{28}^{2+}$ and $AlH^{2+}$ ions on APT mass spectra that may lead to a slight overestimation of the Si concentration, the volume fractions were determined using Mg concentrations (table 1). For both samples, the precipitate volume fraction is similar ($f_v \approx 1.3\%$) showing that there is no significant influence of H. Then, to evaluate if any microstructural features could explain the hardness difference reported in fig. 2, some TEM observations were carried out.

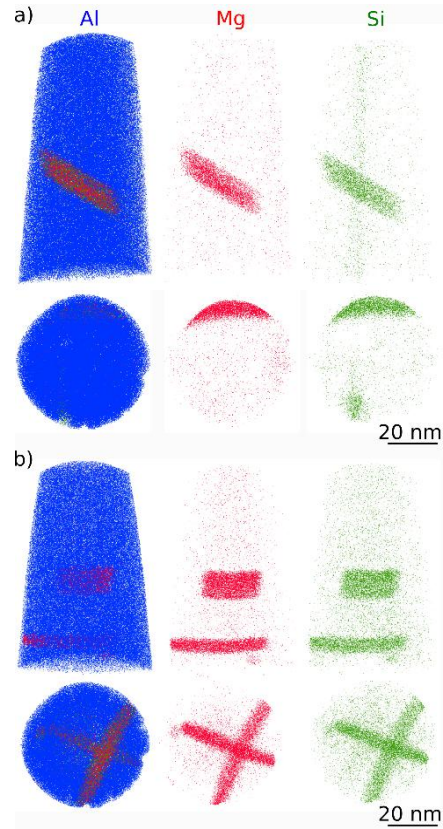

Figure 4: Three dimensional reconstructions of volumes analysed in the AlMgSi alloy aged 24h at 200°C in air (a) and under $H_2$ (b). For each state the same volume is represented three times: with all atoms (Al blue, Mg blue, Si green), only Mg and only Si to clearly exhibit the rod-shaped precipitates.

Table 1: Micro-hardness HV, mean spacing λ between rod shaped precipitates intersecting (001) planes from STEM images, fig. 6), matrix and precipitate compositions (from APT data) and precipitate volume fraction (estimated from eq. (2) and using compositions measured by APT) of the AlMgSi alloy aged 24h at 200°C in air and in $H_2$.

|  | El. | Matrix (at.%) | Precipitate (at.%) | λ (nm) | HV | $f_v(\%)$ |
|---|---|---|---|---|---|---|
| Air | Al | *bal.* | *bal.* |  |  |  |
|  | Mg | 0.01 ± 0.003 | 63.7 ± 1.7 | 148 ± 6 | 76 ± 3 | 1.30 ± 0.08 |
|  | Si | 0.09 ± 0.007 | 29.1 ± 1.6 |  |  |  |
| $H_2$ | Al | *bal.* | *bal.* |  |  |  |
|  | Mg | 0.03 ± 0.006 | 64.7 ± 1.4 | 126 ± 6 | 92 ± 3 | 1.25 ± 0.24 |
|  | Si | 0.12 ± 0.01 | 31.2 ± 1.4 |  |  |  |

At 200°C, in samples aged in air or under $H_2$, TEM bright field images taken in <001> zone axis (Figs. 5.a and 5.b) clearly exhibit classical $\beta'$ rod-shaped precipitates aligned along the three equivalent <100> directions of the FCC lattice of aluminium [43,45–50]. High Resolution TEM (HRTEM) images of these precipitates (Figs. 5c, 5e, 5g, 5i) show that two cross sectional morphologies could be observed in both samples (ellipsoidal or circular). The analyses of structures and relationships with the matrix from fast Fourier transforms of images (Figs. 5d, 5h, 5f, 5j) indicate that they correspond to two types of precipitate (mostly $\beta"$, and B') [43,46,50]. Anyway, in agreement with APT data, precipitates look rather similar in both samples, aged in air or under $H_2$ with a length in a range of 80 to 200 nm and a diameter estimated from the high-resolution images in a range of 5 to 10 nm.

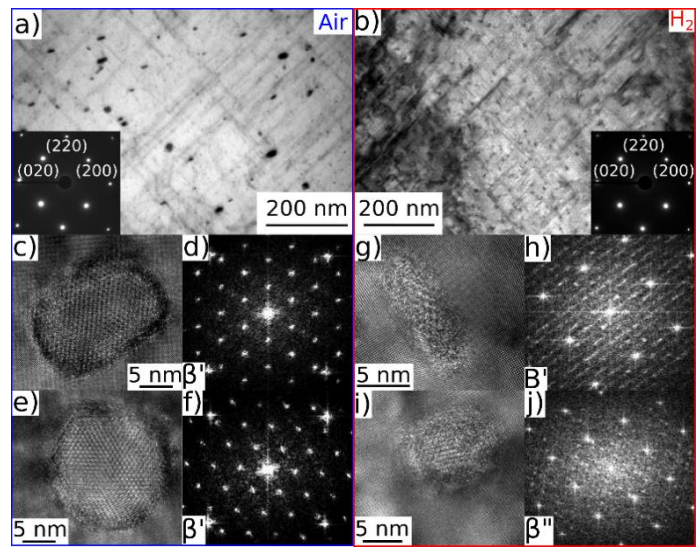

Figure 5: TEM images of the AlMgSi alloy aged 24h at 200°C under air (a,c-d) and $H_2$ (b, e-h). Bright field TEM images (a, b), HRTEM images in <001> zone axis showing the cross section of precipitate (c, e, g, i) with the corresponding FFT patterns (d, h, f, j).

STEM-HAADF images in <001> zone axis (fig. 6) are more reliable to accurately count precipitates parallel to <001> because they only result from a Z-contrast and not from local lattice distortions such as TEM bright field images. Since rod shaped precipitates with the main axis parallel to the electron beam exhibit the best contrast, only these precipitates were considered and counted in large areas to determine the mean distance between precipitates in both samples. In the alloy aged in air, 420 precipitates were counted on a 9.26 $\mu m^2$ area against 519 precipitates on 8.49 $\mu m^2$ in the alloy aged in $H_2$. When a Poisson distribution is considered, then the statistical error is below 5% (4.3 % and 4.8% respectively). Finally, assuming a homogeneous distribution of precipitates, it leads to a mean spacing between precipitates intersecting (001) planes of $\lambda_A = 148\pm6$ nm and of $\lambda_H = 128\pm6$ nm respectively. Precipitates being similar in both samples, the smaller mean spacing indicates that there is a higher precipitate density in the alloy aged under $H_2$ which is consistent with a higher micro-hardness (fig. 2). Since the volume fraction is similar in both alloy (from APT data, table 1), this is obviously the result of a slower precipitate coarsening at 200°C under $H_2$.

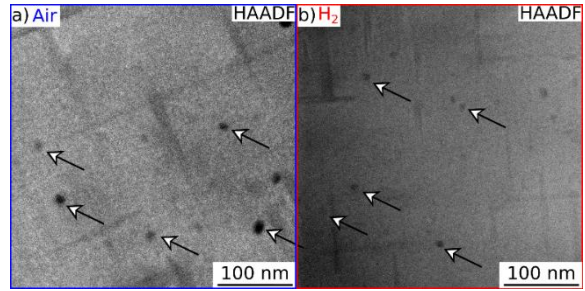

Figure 6: STEM HAADF images (<001> zone axis) of the AlMgSi alloy aged 24h at 200°C under air (a) and $H_2$ (b). Rod-shaped precipitates aligned along the <001> direction and thus parallel to the electron beam are arrowed.

## *4.2 Origin of the highest hardness of the AlMgSi alloy aged at 200°C under $H_2$*

Microstructural characterisation indicates that hydrogen slows down precipitate coarsening leading to a higher mechanical strength of the alloy after AA in $H_2$. The yield stress $\sigma_Y$ of metallic alloys depends on microstructural features and additive contributions may be considered as a first approximation [44,51,52]:

$$\sigma_Y = \sigma_0 + \sigma_d + \sigma_{gb} + \sigma_{ss} + \sigma_P, \qquad (3)$$

where $\sigma_0$ is the friction stress, $\sigma_d$ the forest hardening due to dislocations, $\sigma_{gb}$ contribution of the grain boundaries, $\sigma_{ss}$ the contribution of solid solution and $\sigma_P$ the contribution of precipitates [44,51,52]. Although, the material goes well beyond the onset of plasticity during the micro-hardness tests, it has been shown that the micro-hardness (HV) is typically proportional to the yield stress, so that:

$$\sigma_Y = \frac{HV}{T}, \qquad (4)$$

with $T$ the Tabor factor, a constant equal to $3.5\ MPa^{-1}$ [51,53]. Thus, using eqs. (3) and (4), the micro-hardness difference between the material aged under $H_2$ and in air is $\Delta HV = 16 \pm 6$ (fig. 2, table 1), and corresponds to a yield stress difference of $\Delta\sigma_Y = 56 \pm 21\ MPa$

Considering $\sigma_0$, $\sigma_d$ and $\sigma_{gb}$ being unchanged in eq. (3), the difference could only result from the solid solution or stress contribution of precipitates $\sigma_P$. The latter results from the interaction between precipitates and dislocations. It is well admitted that the mechanisms depend on the precipitate size: smallest precipitates are sheared and beyond a critical radius, they are by-passed [54]. In AlMgSi alloys, it has been shown that the critical radius of rod shaped precipitates is between 2 nm and 5 nm [44,55]. Our TEM observations (fig. 5), clearly show that after ageing at 200°C during 24h in air or under $H_2$, precipitates are beyond this critical size and thus are most likely by-passed. Then, assuming that they are all homogenously by-passed by moving dislocations, their contribution to the yield stress $\sigma_P$ writes as [54,55]:

$$\sigma_P = \frac{2M\beta\mu b}{L^{111}}, \qquad (5)$$

with $L^{111}$ the mean precipitate spacing in (111) planes, $M$ the Taylor factor, $\beta$ a geometric factor, $\mu$ the shear modulus and $b$ the Burger vector of <110> dislocations in (111) planes. These parameters are listed in table 2. From STEM images (fig. 6), the mean spacing $\lambda$ between precipitates intersecting (100) planes was determined in both samples (table 1). Since there are three equivalent (100) planes in the FCC lattice and using geometrical considerations [35], we have:

$$L^{111} = \frac{\sqrt{3}}{3} \times \lambda. \qquad (6)$$

Table 2: parameters of eq. (5), from [36]

| Parameter | Value |
| --- | --- |
| $M$ | 3 |
| $\beta$ | 0.36 |
| $\mu$ | 27 GPa |
| $b$ | 0.284 nm |

Then, from eqs. (5) and (6), $\Delta\sigma_p = \sigma_p^{H2} - \sigma_p^{air} = 32 \pm 22\ MPa$. This stress contribution variation is slightly lower than the yield stress variation estimated from micro-hardness values using eq. (4) ($\Delta\sigma_Y = 56 \pm 21\ MPa$). Even though the error bar of these estimates is relatively large, it cannot be excluded that another hardening contribution increases the hardness of the material aged under $H_2$. Both samples exhibit very similar amount of Mg and Si in solid solution (table 1), but additional solute strengthening might come from hydrogen atoms incorporated during the treatment. However, at 200°C, the equilibrium solubility of hydrogen in pure aluminium is only 1.7 10$^{-7}$ at. [56], thus significant hardening resulting from interactions between hydrogen atoms in solid solution and dislocations is quite unlikely or negligible compared to the hardening contribution of precipitates.

Previous work has shown however that $\beta'$ precipitates/matrix interfaces could be trapping sites for tritium [27]. Hydrogen atoms trapped at these interfaces could increase the stress required to by-pass precipitates leading to an additional increase of the yield stress. Nevertheless, it can be concluded that most of the micro-hardness difference results from a slower kinetic of precipitate coarsening due to hydrogen. The underlying physical mechanisms are discussed in the next section based on classical Oswald ripening theory [57,58].

*4.3 Influence of hydrogen on precipitate ripening*

Precipitate coarsening, also called Oswald ripening, is relatively well described by the classical Lifshitz, Sloyzoz and Wagner (LSW) theories. The original concepts were established for spherical precipitates [57], but they have been extended for various geometries, including rod shaped precipitates in AlMgSi alloys [58,59]. This theory gives the following relationship between the precipitate radius evolution and time:

$$\bar{r}^3 - \bar{r}_0^3 = \frac{\delta\, C_i \Omega_P^2 D_m \gamma}{RT}(t - t_0), \qquad (7)$$

where $\bar{r}^3$ and $\bar{r}_0^3$ are the mean precipitate radius at time $t$ and $t_0$, respectively; $\delta$ a dimensionless constant that depends on the precipitate morphology ($\delta = 4/9$ for cylinders [58,60]); $\gamma$ the free-energy of the matrix/precipitate interface; $T$ the temperature; $R$ the gas constant; $\Omega_p$ the molar volume of the precipitate, $D_m$ the diffusion coefficient of the solute controlling the precipitation kinetic and $C_i$ the concentration of this element at the matrix/precipitate interface. In the present study, precipitates contain twice more Mg than Si (table 1) and at 200°C, the slowest solute is Mg [61], thus the coarsening kinetic should be mainly controlled by the atomic mobility of Mg.

The matrix and precipitate compositions are very similar in samples aged under air and $H_2$ (table 1), thus the reduction of the coarsening kinetic under hydrogen can only be the result of (i) a reduction of the interfacial free-energy $\gamma$, or/and (ii) a reduction of the solute diffusion coefficient $D_m$. Semi-coherent and incoherent precipitate/matrix interfaces with enhanced free volumes or local lattice distortions could be indeed potential trapping sites for hydrogen which may affect the cohesion energy as considered in the HEDE model for HE [7–9,27]. Hydrogen atoms are also known to strongly interact with vacancies [28,41,62], they could affect both their formation and migration energies and thus significantly affect the atomic mobility of Mg and Si in the matrix, as observed during natural ageing.

Assuming that the length $l$ of $\beta'$ rod shaped precipitates is proportional to their radius $\bar{r}$ ($l = \xi \bar{r}$, with $\xi$ a shape factor), other authors have estimated using small-angle neutron scattering data in a similar AlMgSi alloy that $\xi$ is about 13 at 200°C [44]. With the same approximation, then the mean volume $\overline{V_p}$ of rod-shaped precipitates writes as:

$$\overline{V_p} = \pi\, \xi\, \bar{r}^3, \qquad (8)$$

$\lambda$ being the mean distance between precipitates crossing each of the three equivalent (100) planes (table 1), then the volume fraction of precipitates writes as, assuming:

$$f_v = 3\bar{V}_p/\lambda^3. \qquad (9)$$

Thus, the combination of eqs (8) and (9) leads to:

$$\bar{r}^3 = f_v \lambda^3/3\pi\xi. \qquad (10)$$

The precipitate volume fraction being similar when the alloy is aged at 200°C in air or under $H_2$ (table 1), then the relative change in interfacial energy diffusion coefficient product yields from eqs. (7) and (10):

$$\frac{\lambda_H^3 - \lambda_0^3}{\lambda_A^3 - \lambda_0^3} = \frac{D_{Mg}^H \gamma_H}{D_{Mg}^A \gamma_A}, \qquad (11)$$

where $\lambda_0$ is the mean distance between precipitates at the beginning of the coarsening. From eq. (11), $D_{Mg}^H \gamma_H / D_{Mg}^A \gamma_A$ was plotted in fig. 7 as a function of $\lambda_0$. A realistic value of $\lambda_0$ can be obtained from APT data collected in the alloy aged during 24h at 160°C (fig. 3) and it seems typically in range of 50 to 100nm. Then, as shown on fig. 7, it leads to $D_{Mg}^H \gamma_H / D_{Mg}^A \gamma_A$ in a range of 0.4 and 0.6. Such result means that hydrogen induces a variation between 40% and 60% of either the diffusion coefficient of Mg or of the precipitates/matrix interfacial energy.

Considering the precipitate number density and the equilibrium solubility of hydrogen in pure aluminium at 200°C ($C_H^{sol} = 1.7\ 10^{-7}$ [56]), then only about 10 hydrogen atoms are available for each precipitate. Such a small amount might not significantly change the interfacial energy. Besides, phase field simulations have shown that a change in interfacial energy affects the morphology of rod-shape precipitates in AlMgSi alloys [63]. From our TEM and APT data (figs. 4, 5 and 6), no significant morphology differences could be observed, indicating that hydrogen most probably does not affect significantly the precipitate/matrix interfacial energy. Then, like during natural ageing, most of differences are the result from a reduction of the diffusion coefficient due to interactions between H atoms and vacancies.

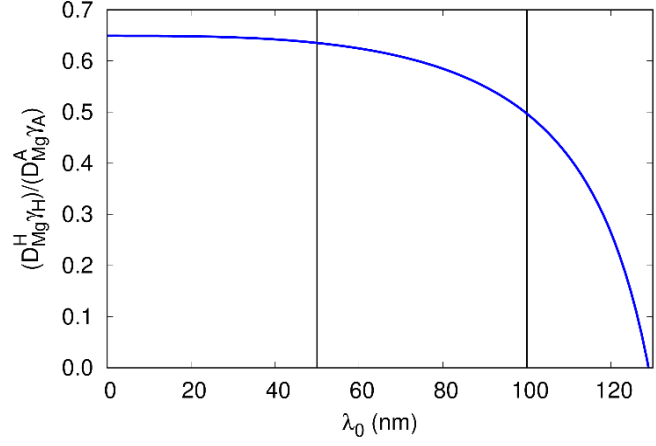

Figure 7: Ratio of diffusion coefficient interfacial energy product with and without H (estimated from eq. (11)) as a function of the mean distance between precipitates intersecting (001) atomic planes at the beginning of the coarsening process ($\lambda_0$). Black lines indicate the range of realistic values for $\lambda_0$ estimated from APT data.

The diffusion coefficient $D_j^i$ is classically linked to the temperature such as:

$$D_j^i = D_0^i \exp\left(-\frac{Q^i}{RT}\right), \qquad (12)$$

Mg being the slowest solute, the coarsening kinetic is controlled by its mobility and the diffusion coefficient of this element will be considered ($D_0^A = 2.2 \times 10^{-4}$ m$^2$.s$^{-1}$, and $Q^A = 130$ kJ.mol$^{-1}$ [44,55,64]). Then, in fig. 2, the micro-hardness after ageing in air during 24h at ~187°C being similar to that under H$_2$ at 200°C, hence $D_{Mg}^H(T = 200°C) \approx D_{Mg}^A(T = 187°C)$. Since $D_{Mg}^A(T = 187°C)/D_{Mg}^A(T = 200°C) \approx 0.4$, it can be concluded that H in solution leads to a decrease of about 60% of the mobility of Mg atoms. This estimate is similar to the estimate done from natural ageing data and similarly this atomic mobility reduction can either be due to an increase of the vacancy formation energy or/and vacancy migration energy [38]. In both artificial (24h at 200°C) and natural ageing, we have more vacancies than hydrogen atoms (at T=200°C, $C_H^{sol} \approx C_{Vac}^{eq}$ in pure Al [56,65] and when the alloy is quenched and aged naturally, vacancies are in sursaturation), thus the effect of hydrogen on the vacancy concentration should be negligible but it may increase the vacancy migration energy. Following the Perks model describing the diffusivity of solute with the vacancy migration energy in the matrix [66,67], the pre-exponential factor and the activation energy of eq. (12) are equivalent to a pre-exponential factor $d_0^i$ and the vacancy migration energy $E_m^{Vac,i}$ in the matrix aged in air or in H$_2$ environment (from the literature, $E_m^{Vac,A} = 0.60$ eV [68]). While the model has its limitation as shown in the literature [67], we can estimate the variation of vacancy migration energy from the variation of diffusion coefficient $D_{eff,S}^i$, which leads to:

$$E_m^{Vac,H} = E_m^{Vac,A} - kT\ln(\Delta), \qquad (13)$$

with $k$ the Boltzmann constant and $\Delta$ the ratio $D_{eff,S}^H/D_{eff,S}^A$ (*i.e*: $\Delta = 0.5$ during natural and artificial ageing at 200°C during 24h). These variations are equivalent to a reduction of 0.02 eV of the vacancy migration energy.

Since we observe a reduction of the solute mobility when the alloy is aged naturally or artificially at 200°C, a reduction of the solute mobility should also be observed when the alloy is aged at 160°C. This reduction of mobility should delay the growth and coarsening of precipitates, inducing a reduction of hardness, which is not observed in figs. 2 and 3. However, at 160°C, the equilibrium concentration of vacancies $C_{Vac}^{eq}$ is twice lower than the hydrogen solubility $C_H^{sol}$ in pure aluminium while the ratio is close to 1 at 200°C [32,56]. In this situation, there are more hydrogen than vacancies and they may promote the formation of additional vacancies [41] (as reported also in pure nickel [62]), in agreement with the SAV model [41]. This enhancement may then compensate the increase of the vacancy migration energy, inducing a minor effect on the growth of the precipitate.

## 5. Conclusions

At all stages of precipitation (nucleation, growth and coarsening), crystalline defects play an important role. More specially, the vacancy concentration and mobility tune the kinetic and interfaces the thermodynamics through the nucleation barrier, precipitate morphology and driving force for coarsening. In this study we have investigated the influence of hydrogen on natural aging and precipitate coarsening in an AlMgSi alloy. This element is indeed known to strongly interact with crystalline defects and the aim was to quantify these interactions. During natural aging, microhardness measurements clearly showed that the hardening response is delayed when the alloy is in NaOH as compared to natural aging in air at the same temperature. It has been attributed to a slower recovery of as quenched non-equilibrium vacancies resulting from hydrogen-vacancy interactions. When the alloy is removed from NaOH, the microhardness quickly catches up that of the material aged in air, confirming that the original vacancy mobility is recovered when H atoms are desorbed. Based on the hardening delay, hydrogen atoms induced a reduction of the mean diffusion coefficient of vacancy by a factor of two.

Since natural ageing occurs in a situation where the vacancy concentration is not constant and not at the equilibrium, further experiments were conducted in $H_2$ environment at higher temperature (160°C and 200°C) where the vacancy concentration is constant and precipitation occur. APT and TEM experimental data show that phase composition and solubility limits of Si and Mg in FCC Al are not affected by H, but a significant reduction of the coarsening of rod-shaped precipitates could be quantified at 200°C. These observations are consistent with micro-hardness measurements showing a larger softening of the material over-aged during 24h at 200°C in air as compared to the alloy aged in similar conditions in $H_2$. Using a simple model to account for all strengthening contributions, the higher micro-hardness of the material aged under $H_2$ is mainly attributed to a higher precipitate density. Then, classical Oswald ripening theories were considered to understand the underlying mechanisms. From this approach it was concluded that the delayed precipitate coarsening only results from a reduced solute diffusion coefficient due to a higher vacancy migration energy. Our estimates indicate that for a concentration of hydrogen close to the vacancy concentration, it increases modestly from 0.6 to only 0.62 eV with a significant impact on long term ageing.

**Acknowledgements**

The authors thanks J.B. Maillet and Prof. H. Zapolsky for fruitful discussions. Dr. L. Perrière and Dr. F. Couturas from ICMPE are gratefully acknowledge for AA carried out under $H_2$ atmosphere. The authors acknowledge the financial support of the French Agence Nationale de la Recherche LabEx EMC3 through the Project SolHyDef (Grant No. 10-LABX-0009), and the Normandy Region (Réseau d'Intérêt Normand - Label d'excellence).

**References**

[1]　R.P. Gangloff, B.P. Somerday, Gaseous Hydrogen Embrittlement of Materials in Energy Technologies, Woodhead Publisher, 2011.

[2]　F. Martin, X. Feaugas, A. Oudriss, D. Tanguy, L. Briottet, J. Kittel, State of hydrogen in matter: Fundamental ad/absorption, trapping and transport mechanisms, in: C. Blanc, I. Aubert (Eds.), Mech. - Microstruct. - Corros. Couplings, Elsevier, 2019: pp. 171--197.

[3]　S.P. Lynch, Hydrogen embrittlement phenomena and mechanisms, Corros. Rev. 30 (2012) 105–123. https://doi.org/10.1515/corrrev-2012-0502.

[4]　J.R. Scully, G.A. Young, S.W. Smith, Hydrogen embrittlement of aluminum and aluminum-based alloys, in: R.P. Gangloff, B.P. Somerday (Eds.), Gaseous Hydrog. Embrittlement Mater. Energy Technol., Woodhead Publisher, 2012: pp. 707–768. https://doi.org/10.1533/9780857093899.3.707.

[5]　R. Kirchheim, A. Pundt, Hydrogen in Metals, in: D.E. Laughlin, K. Hono (Eds.), Phys. Metall., 5th ed., Elsevier, Oxford, 2014: pp. 2597–2705. https://doi.org/10.1016/B978-0-444-53770-6.00025-3.

[6]　I.M. Robertson, P. Sofronis, A. Nagao, M.L. Martin, S. Wang, D.W. Gross, K.E. Nygren, Hydrogen Embrittlement Understood, Metall. Mater. Trans. A. 46 (2015) 2323–2341. https://doi.org/10.1007/s11661-015-2836-1.

[7]　A.R. Troiano, The Role of Hydrogen and Other Interstitials in the Mechanical Behavior of Metals, Trans. Am. Soc. Met. 52 (1960) 54–60. https://doi.org/10.1007/s13632-016-0319-4.

[8]　R.A. Oriani, The diffusion and trapping of hydrogen in steel, Acta Metall. 18 (1970) 147–157. https://doi.org/10.1016/0001-6160(70)90078-7.

[9]　S.P. Lynch, Hydrogen embrittlement and liquid-metal embrittlement in nickel single crystals, Scr. Metall. 13 (1979) 1051–1056. https://doi.org/10.1016/0036-9748(79)90202-3.

[10]　C.D. Beachem, A New Model for Hydrogen-Assisted Cracking ( Hydrogen " Embrittlement "), Metall. Trans. 3 (1972) 437–451.

[11]  H.K. Birnbaum, P. Sofronis, Hydrogen-enhanced localized plasticity--a mechanism for hydrogen-related fracture, Mater. Sci. Eng. A. 176 (1994) 191–202.

[12]  P. Sofronis, Y. Liang, N. Aravas, Hydrogen induced shear localization of the plastic flow in metals and alloys, Eur. J. Mech. A/Solids. 20 (2001) 857–872. https://doi.org/10.1016/S0997-7538(01)01179-2.

[13]  D. Delafosse, T. Magnin, Hydrogen induced plasticity in stress corrosion cracking of engineering systems, Eng. Fract. Mech. 68 (2001) 693–729. https://doi.org/10.1016/S0013-7944(00)00121-1.

[14]  R.B. McLellan, Z.R. Xu, Hydrogen-induced vacancies in the iron lattice, Scr. Mater. 36 (1997) 1201–1205. https://doi.org/10.1016/S1359-6462(97)00015-8.

[15]  M. Nagumo, Function of hydrogen in embrittlement of high-strength steels, ISIJ Int. 41 (2001) 590–598.

[16]  Y. Fukai, Formation of superabundant vacancies in M-H alloys and some of its consequences: A review, J. Alloys Compd. 356–357 (2003) 263–269. https://doi.org/10.1016/S0925-8388(02)01269-0.

[17]  N.Z. Carr, R.B. McLellan, The thermodynamic and kinetic behavior of metal-vacancy-hydrogen systems, Acta Mater. 52 (2004) 3273–3293. https://doi.org/10.1016/j.actamat.2004.03.024.

[18]  E.P. Georgiou, J.P. Celis, C.N. Panagopoulos, The effect of cold rolling on the hydrogen susceptibility of 5083 aluminum alloy, Metals (Basel). 7 (2017). https://doi.org/10.3390/met7110451.

[19]  R. Kirchheim, Reducing grain boundary, dislocation line and vacancy formation energies by solute segregation. I. Theoretical background, Acta Mater. 55 (2007) 5129–5138. https://doi.org/10.1016/j.actamat.2007.05.047.

[20]  H. Khanchandani, S.-H. Kim, R.S. Varanasi, T.S. Prithiv, L.T. Stephenson, B. Gault, Hydrogen and deuterium charging of site-specific specimen for atom probe tomography, 2021.

[21]  Y.S. Chen, H. Lu, J. Liang, A. Rosenthal, H. Liu, G. Sneddon, I. McCarroll, Z. Zhao, W. Li, A. Guo, J.M. Cairney, Observation of hydrogen trapping at dislocations, grain boundaries, and precipitates, Science (80-. ). 367 (2020) 171–175. https://doi.org/10.1126/science.aaz0122.

[22]  C.R. Hutchinson, Modeling the kinetics of precipitation in aluminium alloys, Woodhead Publishing Limited, 2010. https://doi.org/10.1533/9780857090256.2.422.

[23]  J.W. Martin, Precipitation Hardening: Theory and Applications, Elsevier Science, 2012.

[24]  A. Deschamps, C.R. Hutchinson, Precipitation kinetics in metallic alloys: Experiments and modeling, Acta Mater. 220 (2021) 117338. https://doi.org/10.1016/j.actamat.2021.117338.

[25]  K. Horikawa, T. Matsubara, H. Kobayashi, Hydrogen charging of Al–Mg–Si-based alloys by friction in water and its effect on tensile properties, Mater. Sci. Eng. A. 764 (2019) 138199. https://doi.org/10.1016/j.msea.2019.138199.

[26]  P. Dumitraschkewitz, P.J. Uggowitzer, S.S.A. Gerstl, J.F. Löffler, S. Pogatscher, Size-dependent diffusion controls natural aging in aluminium alloys, Nat. Commun. 10 (2019). https://doi.org/10.1038/s41467-019-12762-w.

[27]  H. Saitoh, Y. Iijima, K. Hirano, Behaviour of hydrogen in pure aluminium, Al-4 mass% Cu and Al-1 mass% Mg2Si alloys studied by tritium electron microautoradiography, J. Mater. Sci. 29 (1994) 5739–5744. https://doi.org/10.1007/BF00349974.

[28]  C. Wolverton, V. Ozoliņš, M. Asta, Hydrogen in aluminum: First-principles calculations of structure and thermodynamics, Phys. Rev. B. 69 (2004) 1–16. https://doi.org/10.1103/PhysRevB.69.144109.

[29]  R. Nazarov, T. Hickel, J. Neugebauer, Ab initio study of H-vacancy interactions in fcc metals: Implications for the formation of superabundant vacancies, Phys. Rev. B - Condens. Matter Mater. Phys. 89 (2014) 1–18.

https://doi.org/10.1103/PhysRevB.89.144108.

[30]  D. Connétable, M. David, Study of vacancy-(H,B,C,N,O) clusters in Al using DFT and statistical approaches: Consequences on solubility of solutes, J. Alloys Compd. 748 (2018) 12–25. https://doi.org/10.1016/j.jallcom.2018.03.081.

[31]  H.K. Birnbaum, C. Buckley, F. Zeides, E. Sirois, P. Rozenak, S. Spooner, J.S. Lin, Hydrogen in aluminum, J. Alloys Compd. 253–254 (1997) 260–264. https://doi.org/10.1016/S0925-8388(96)02968-4.

[32]  H. Saitoh, Y. Iijima, H. Tanaka, hydrogen diffusivity in aluminim measured by a glow discharge permeation method, Acta Metall. Mater. 42 (1994) 2493--2498. https://doi.org/10.1016/0956-7151(94)90329-8.

[33]  M.K. Miller, A. Cerezo, H.G. Hetherington, G.D.W. Smith, Atom probe field ion microscopy, Oxford Science Publications -- Clarendon Press, 1996.

[34]  T. Masuda, X. Sauvage, S. Hirosawa, Z. Horita, Achieving highly strengthened Al–Cu–Mg alloy by grain refinement and grain boundary segregation, Mater. Sci. Eng. A. 793 (2020) 139668. https://doi.org/10.1016/j.msea.2020.139668.

[35]  J. Banhart, C.S.T. Chang, Z. Liang, N. Wanderka, M.D.H. Lay, A.J. Hill, Natural aging in Al-Mg-Si alloys - A process of unexpected complexity, Adv. Eng. Mater. 12 (2010) 559–571. https://doi.org/10.1002/adem.201000041.

[36]  J. Banhart, M.D.H. Lay, C.S.T. Chang, A.J. Hill, Kinetics of natural aging in Al-Mg-Si alloys studied by positron annihilation lifetime spectroscopy, Phys. Rev. B - Condens. Matter Mater. Phys. 83 (2011) 1–13. https://doi.org/10.1103/PhysRevB.83.014101.

[37]  Y. Weng, L. Ding, Y. Xu, Z. Jia, Q. Sun, F. Chen, X. Sun, Y. Cong, Q. Liu, Effect of In addition on the precipitation behavior and mechanical property for Al-Mg-Si alloys, J. Alloys Compd. 895 (2022) 162685. https://doi.org/10.1016/j.jallcom.2021.162685.

[38]  M. Mantina, Y. Wang, R. Arroyave, L.Q. Chen, Z.K. Liu, C. Wolverton, First-principles calculation of self-diffusion coefficients, Phys. Rev. Lett. 100 (2008) 1–4. https://doi.org/10.1103/PhysRevLett.100.215901.

[39]  Y. Wang, D. Connétable, D. Tanguy, Hydrogen influence on diffusion in nickel from first-principles calculations, Phys. Rev. B. 91 (2015). https://doi.org/10.1103/PhysRevB.91.094106.

[40]  J.P. Du, W.T. Geng, K. Arakawa, J. Li, S. Ogata, Hydrogen-Enhanced Vacancy Diffusion in Metals, J. Phys. Chem. Lett. 11 (2020) 7015–7020. https://doi.org/10.1021/acs.jpclett.0c01798.

[41]  Y. Fukai, The Metal-Hydrogen System, Springer Berlin Heidelberg, 2006.

[42]  M. Yang, A. Orekhov, Z.Y. Hu, M. Feng, S. Jin, G. Sha, K. Li, V. Samaee, M. Song, Y. Du, G. Van Tendeloo, D. Schryvers, Shearing and rotation of β″ and β′ precipitates in an Al-Mg-Si alloy under tensile deformation: In-situ and ex-situ studies, Acta Mater. 220 (2021) 117310. https://doi.org/10.1016/j.actamat.2021.117310.

[43]  C.D. Marioara, H. Nordmark, S.J. Andersen, R. Holmestad, Post-β″ phases and their influence on microstructure and hardness in 6xxx Al-Mg-Si alloys, J. Mater. Sci. 41 (2006) 471–478. https://doi.org/10.1007/s10853-005-2470-1.

[44]  D. Bardel, M. Perez, D. Nelias, A. Deschamps, C.R. Hutchinson, D. Maisonnette, T. Chaise, J. Garnier, F. Bourlier, Coupled precipitation and yield strength modelling for non-isothermal treatments of a 6061 aluminium alloy, Acta Mater. 62 (2014) 129–140. https://doi.org/10.1016/j.actamat.2013.09.041.

[45]  G.A. Edwards, K. Stiller, G.L. Dunlop, M.J. Couper, The precipitation sequence in Al-Mg-Si alloys, Acta Mater. 46 (1998) 3893–3904. https://doi.org/10.1016/S1359-6454(98)00059-7.

[46]   J.Y. Yao, D.A. Graham, B. Rinderer, M.J. Couper, A TEM study of precipitation in Al-Mg-Si alloys, Micron. 32 (2001) 865–870. https://doi.org/10.1016/S0968-4328(00)00095-0.

[47]   M.H. Jacobs, Precipitates formed during ageing of an Al-Mg-Si alloy, Philos. Mag. 26 (1972) 1–13. https://doi.org/10.1051/jphystap:018930020027701.

[48]   G. Meyruey, V. Massardier, W. Lefebvre, M. Perez, Over-ageing of an Al-Mg-Si alloy with silicon excess, Mater. Sci. Eng. A. 730 (2018) 92–105. https://doi.org/10.1016/j.msea.2018.05.094.

[49]   G. Lu, B. Sun, J. Wang, Y. Liu, C. Liu, High-temperature age-hardening behavior of Al–Mg–Si alloys with varying Sn contents, J. Mater. Res. Technol. 14 (2021) 2165–2173. https://doi.org/10.1016/j.jmrt.2021.07.122.

[50]   Y. Liu, Y.X. Lai, Z.Q. Chen, S.L. Chen, P. Gao, J.H. Chen, Formation of β"-related composite precipitates in relation to enhanced thermal stability of Sc-alloyed Al-Mg-Si alloys, J. Alloys Compd. 885 (2021) 160942. https://doi.org/10.1016/j.jallcom.2021.160942.

[51]   A. Rodríguez-Veiga, B. Bellón, I. Papadimitriou, G. Esteban-Manzanares, I. Sabirov, J. LLorca, A multidisciplinary approach to study precipitation kinetics and hardening in an Al-4Cu (wt. %) alloy, J. Alloys Compd. 757 (2018) 504–519. https://doi.org/10.1016/j.jallcom.2018.04.284.

[52]   R.H. Wang, Y. Wen, B.A. Chen, Sn microalloying Al–Cu alloys with enhanced fracture toughness, Mater. Sci. Eng. A. 814 (2021) 141243. https://doi.org/10.1016/j.msea.2021.141243.

[53]   I.M. Hutchings, The contributions of David Tabor to the science of indentation hardness, J. Mater. Res. 24 (2009) 581–589. https://doi.org/10.1557/jmr.2009.0085.

[54]   A. Deschamps, Y. Brechet, Influence of predeformation and ageing of an Al-Zn-Mg Alloy-II. Modeling of precipitation kinetics and yield stress, Acta Mater. 47 (1998) 293–305. https://doi.org/10.1016/S1359-6454(98)00296-1.

[55]   O.R. Myhr, O. Grong, S.J. Andersen, Modelling of the age hardening behaviour of Al-Mg-Si alloys, Acta Mater. 49 (2001) 65–75. https://doi.org/10.1016/S1359-6454(00)00301-3.

[56]   A. San-Martin, F.D. Manchester, The Al-H (aluminum-hydrogen) system, J. Phase Equilibria. 13 (1992) 17–21. https://doi.org/10.1007/BF02645371.

[57]   A. Baldan, Progress in Ostwald ripening theories and their applications in nickel-base super alloys, J. Mater. Sci. 37 (2002) 2379–2405.

[58]   M. Usta, M.E. Glicksman, R.N. Wright, The effect of heat treatment on Mg2Si coarsening in aluminum 6105 alloy, Metall. Mater. Trans. A Phys. Metall. Mater. Sci. 35 A (2004) 435–438. https://doi.org/10.1007/s11661-004-0354-7.

[59]   M. McLean, Microstructural instabilities in metallurgical systems—a review, Met. Sci. 12 (1978) 113–122. https://doi.org/10.1179/msc.1978.12.3.113.

[60]   K. Kim, P.W. Voorhees, Ostwald ripening of spheroidal particles in multicomponent alloys, Acta Mater. 152 (2018) 327–337. https://doi.org/10.1016/j.actamat.2018.04.041.

[61]   L.F. Mondolfo, Aluminium alloys: structure and properties, Buttherworths, London, 1979.

[62]   G. Hachet, A. Metsue, A. Oudriss, X. Feaugas, Influence of hydrogen on the elastic properties of nickel single crystal: A numerical and experimental investigation, Acta Mater. 148 (2018). https://doi.org/10.1016/j.actamat.2018.01.056.

[63]   H. Mao, Y. Kong, D. Cai, M. Yang, Y. Peng, Y. Zeng, G. Zhang, X. Shuai, Q. Huang, K. Li, H. Zapolsky, Y. Du, β'' needle-shape precipitate formation in Al-Mg-Si alloy: Phase field simulation and experimental verification,

Comput. Mater. Sci. 184 (2020). https://doi.org/10.1016/j.commatsci.2020.109878.

[64] P.I. Sarafoglou, A. Serafeim, I.A. Fanikos, J.S. Aristeidakis, G.N. Haidemenopoulos, Modeling of microsegregation and homogenization of 6xxx Al-alloys including precipitation and strengthening during homogenization cooling, Materials (Basel). 12 (2019). https://doi.org/10.3390/ma12091421.

[65] S. Nenno, J.W. Kauffman, Nenno1960.pdf, J. Phys. Soc. Japan. 15 (1960) 220--226. https://doi.org/10.1143/JPSJ.15.220.

[66] J.M. Perks, A.D. Marwick, C.A. English, A computer code to calculate radiation-induced segregation in concentrated ternary alloys, in: AERE Rep. 12121, UKAEA Atomic Energy Research Esthablishement, 1986.

[67] T.R. Allen, G.S. Was, Modeling radiation-induced segregation in austenitic Fe-Cr-Ni alloys, Acta Mater. 46 (1998) 3679–3691. https://doi.org/10.1016/S1359-6454(98)00019-6.

[68] G. Ho, M.T. Ong, K.J. Caspersen, E.A. Carter, Energetics and kinetics of vacancy diffusion and aggregation in shocked aluminium via orbital-free density functional theory, Phys. Chem. Chem. Phys. 9 (2007) 4951–4966. https://doi.org/10.1039/b705455f.